\documentclass[preprint,aps,12pt,preprintnumbers,eqsecnum,nofootinbib]{revtex4}
\usepackage{graphicx}
\usepackage{subfigure}

\usepackage{color}
\usepackage{amssymb,amsmath,amsfonts}
\usepackage{epstopdf} 
\usepackage{braket}

\newcommand{\beq}{\begin{equation}}
\newcommand{\enq}{\end{equation}}

\renewcommand{\ol}{\overline}
\newcommand{\bk}{\braket}
\newcommand{\x}{\times}
\newcommand{\bd}[1]{{\bf #1}}
\newcommand{\cd}{\cdot}
\newcommand{\da}{\dagger}
\newcommand{\Tr}{\text{Tr}}

\begin{document}
%
%
\title{\vspace*{0.5in} 
Flavor from the double tetrahedral group without supersymmetry
\vskip 0.1in}
\author{Christopher D. Carone}\email[]{cdcaro@wm.edu}
\author{Shikha Chaurasia}\email[]{scchaurasia@email.wm.edu}
\author{Savannah Vasquez}\email[]{svasquez@mail.usf.edu}

\affiliation{High Energy Theory Group, Department of Physics,
College of William and Mary, Williamsburg, VA 23187-8795}

\date{November 2, 2016}

\begin{abstract}
We consider a class of flavor models proposed by Aranda, Carone and Lebed, relaxing the assumption of supersymmetry
and allowing the flavor scale to float anywhere between the weak and Planck scales.   We perform global fits to the charged fermion masses 
and CKM angles, and consider the dependence of the results on the unknown mass scale of the flavor sector.  We find that the typical 
Yukawa textures in these models provide a good description of the data over a wide range of flavor scales, with a preference for those
that approach the lower bounds allowed by flavor-changing-neutral-current constraints.  Nevertheless, the possibility that the flavor scale and 
Planck scale are identified remains viable.  We present models that demonstrate how the assumed textures can arise most simply in a 
non-supersymmetric framework.
\end{abstract}
\maketitle

\section{Introduction} \label{sec:intro}

There is a vast literature on models that attempt to explain the observed hierarchy of fermion masses by means of horizontal symmetries.  In
this paper, we revisit one such model, proposed by Aranda, Carone and Lebed, based on the double tetrahedral group 
$T'$~\cite{Aranda:1999kc,Aranda:2000tm}.  Prior to this work, it had been shown that supersymmetric grand unified theories with U(2) 
flavor symmetry predict simple forms for the Yukawa matrices, ones that provide a successful description of charged fermion masses and 
the Cabibbo-Kobayashi-Maskawa (CKM) mixing matrix~\cite{Barbieri:1995uv,Barbieri:1996ww}.  The authors of Ref.~\cite{Aranda:1999kc,Aranda:2000tm} 
posed a simple question:  What is the smallest discrete flavor group that predicts the same form for the Yukawa textures?  The answer to this question was 
determined by the specific group theoretic properties of U(2) that were utilized in the most successful U(2) models~\cite{Barbieri:1996ww}: 
\begin{enumerate}
\item U(2) models involved fields in ${\bf 1}$, ${\bf 2}$ and ${\bf 3}$ dimensional representations (reps).  Matter fields of the 
three generations were embedded into ${\bf 2}\oplus{\bf 1}$ dimensional reps; the fact that the third generation fields were treated differently allowed the model 
to accommodate an order one ({\em i.e.}, a flavor-group-invariant) top quark Yukawa coupling.   The flavor-symmetry-breaking fields, called flavons, appeared in 
all three of these representations. 
\item  In each Yukawa matrix, the two-by-two block associated with the first two generations decomposed into an antisymmetric and symmetric part.  These
followed from the couplings of the ${\bf 1}$ and ${\bf 3}$-dimensional flavon fields, respectively, due to the group multiplication rule 
\begin{equation} 
{\bf 2} \otimes {\bf 2} = {\bf 3} \oplus {\bf 1} \,\,\, . 
\end{equation}
\item   The U(2) symmetry was broken to a U(1) subgroup that rotated all first generation fields by a phase.   This U(1) symmetry was subsequently broken 
at a lower energy scale than that of the original U(2) symmetry.  Since Yukawa couplings emerge as a ratio of a symmetry-breaking scale to a cut off, the sequential
breaking of the flavor symmetry explains why the Yukawa couplings associated with first generation were smaller than those of the heavier generations.
\end{enumerate}
The group $T'$ is special in that it is the smallest discrete group that has ${\bf 1}$, ${\bf 2}$ and ${\bf 3}$-dimensional representations, as well as the multiplication
rule ${\bf 2} \otimes {\bf 2} = {\bf 3} \oplus {\bf 1}$.   We will briefly review the representations and multiplication rules for $T'$ symmetry in Sec.~\ref{sec:textures}.   Following 
Ref~\cite{Aranda:1999kc,Aranda:2000tm}, the appropriate symmetry breaking sequence is achieved if the flavor group includes an Abelian factor, so that $G_F = T' \times Z_3$.   Then the 
breaking pattern of the U(2) model
\begin{equation}
U(2) \stackrel{\epsilon}{\longrightarrow} U(1) \stackrel{\epsilon'}{\longrightarrow} \mbox{ nothing},
\label{eq:ubreak}
\end{equation}
is mimicked by
\begin{equation}
T' \times Z_3 \stackrel{\epsilon}{\longrightarrow} Z^D_3 \stackrel{\epsilon'}{\longrightarrow} \mbox{ nothing}.
\label{eq:tpbreak}
\end{equation}
Here we have indicated the scale of each symmetry breaking via the dimensionless parameters $\epsilon$ and $\epsilon'$, which represent the 
ratio of a symmetry-breaking vacuum expectation value (vev) to the cut off of the effective theory.  We refer to the cut off as the flavor scale, $M_F$, henceforth.   A useful 
way to understand the connection between Eq.~(\ref{eq:ubreak}) and (\ref{eq:tpbreak}) is to consider the SU(2)$\times$U(1) subgroup of U(2);  
The $T'$ factor is a subgroup of the SU(2) factor while $Z_3$ is a subgroup of the U(1).   The $Z_3$ factor remaining after the first step in 
the symmetry-breaking chain in Eq.~(\ref{eq:tpbreak}) also transforms all first generation fields by a phase and will be specified later. The $T' \times Z_3$ model defined in 
this way reproduces the successful Yukawa textures of the U(2) models, but with a much smaller symmetry group.  For other productive applications of $T'$ symmetry in
flavor model building, we refer the reader to Ref.~\cite{manytp}.

The $T'$ models of Refs.~\cite{Aranda:1999kc,Aranda:2000tm} were constructed more than 16 years ago, when it was widely assumed 
that weak-scale supersymmetry was the likely solution to the gauge hierarchy problem.  The numerical study of the Yukawa textures 
in these references assumed supersymmetric renormalization group equations to relate the predictions of the theory at the flavor scale 
$M_F$ to those at observable energies.  Superpartners were taken to have masses just above the electroweak scale, while $M_F$ 
was identified with the scale of supersymmetric grand unification, $\sim 2 \times 10^{16}$~GeV.  The latter choice was motivated by the 
most elegant $T'$  models, which were formulated in the context of an $SU(5)$ grand unified theory.  Some of the essential features of the Yukawa textures 
followed from the combined restrictions of the flavor and grand unified symmetries.   

At the present moment, however, the status of weak-scale supersymmetry as a necessary ingredient in model building is far less certain.  The latest
data from the LHC has found no evidence for supersymmetry.  Of course, this may simply mean that the scale of the superpartner
masses is slightly higher than what one might prefer from the perspective of naturalness; this interpretation would have little
effect on the results of Refs.~\cite{Aranda:1999kc,Aranda:2000tm}.  On the other hand, the LHC may be hinting that there is no 
necessary connection between the weak scale and the scale of supersymmetry breaking.  In this case, one might entertain the possibility
that the supersymmetry breaking scale is associated with the only higher physical mass scale whose existence is well established: the Planck 
scale.  For example, it has been suggested in Ref.~\cite{Ibe:2013rpa} that the shallowness of the Higgs potential may be explained by Planck-scale supersymmetry 
breaking, assuming that supersymmetry is still relevant for a quantum gravitational completion.  This latter assumption itself
has been challenged in Ref.~\cite{Abel:2015oxa}, where it has been noted that there are consistent string theories that are fundamentally 
non-supersymmetric and whose low-energy limit could include the standard model.  Whether supersymmetry is broken at the Planck scale, or not 
present at any scale, one might attempt to address the hierarchy between the weak scale and Planck scale, for example,
by anthropic selection, or by Higgs field relaxation~\cite{Graham:2015cka}, or by mechanisms not yet known.   Alternatively, one might pursue the idea
that quantum gravitational physics does not contribute to scalar field quadratic divergences in the way that one expects naively 
from effective field theory arguments~\cite{Dubovsky:2013ira}.  In this paper, we remain completely agnostic on the issue of naturalness. We 
instead investigate a question that can be addressed in a more definitive and quantitative way:  how well do the $T'$ flavor models in 
Refs.~\cite{Aranda:1999kc,Aranda:2000tm} work if there is no supersymmetry below the Planck scale?

We begin our study by assuming a standard form for the Yukawa textures expected in models with $T' \times Z_3$ symmetry and perform
a global fit to the charged fermion masses and CKM elements assuming that the predictions at the flavor scale $M_F$ are related to those
at the weak scale via non-supersymmetric renormalization group equations\footnote{Note that we do not consider neutrino physics in the present work due to the 
additional model dependence affecting that sector of the theory.   For example, the structure of the theory is different depending on whether neutrino masses are 
Dirac or Majorana, whether the Majorana masses arise via a seesaw mechanism or via coupling to electroweak triplet Higgs fields, and whether additional 
neutral fermions are present with which the neutrinos can mix. We reserve such a study for future work.}.  In the absence of supersymmetry, we no longer have
gauge coupling unification and therefore do not consider grand unified embeddings.   The flavor scale is taken as a free parameter that
may vary anywhere from the TeV scale to the Planck scale.   By study of the goodness of these fits, we consider whether there is any 
preference for a higher or lower flavor scale within the specified range.  If one were to find acceptable results for values of $M_F$ near the Planck scale, 
one might conclude that the model is consistent with a minimal scenario in which there are no other energy scales of physical relevance other than the weak 
and the Planck scale.  On the other hand, if one were to find acceptable results for $M_F$ closer to the lower bounds from flavor-changing-neutral-current
processes, then one might obtain interesting predictions for observable indirect effects of heavy particles associated with the flavor sector.  

Our paper is organized as follows.  In the next section, we briefly review the flavor models of interest and present a parameterization of
the Yukawa matrix textures that typically arise in these models at the flavor scale $M_F$.  In Sec.~\ref{sec:numerical}, we study
the predictions that follow from these textures by a non-supersymmetric renormalization group analysis, including global fits to the current data on charged fermion 
masses and CKM elements.   In Sec.~\ref{sec:fcnc}, we point out the largest indirect effects of heavy 
flavor-sector particles on flavor-changing-neutral current processes in the case where $M_F$ is low.  In Sec.~\ref{sec:models}, we address model building issues:  
supersymmetric models have two Higgs doublets  (in order to cancel anomalies) and have a superpotential that is constrained by holomorphicity; these requirements 
are absent in the non-supersymmetric case.  Hence, in this section we show how the textures assumed in Sec.~\ref{sec:numerical} may arise in 
non-supersymmetric $T'$ models.   In the final section, we summarize our conclusions.

\section{Typical Yukawa textures from T-prime symmetry} \label{sec:textures}

The group $T'$ is discussed at length in Ref.~\cite{Aranda:2000tm}.  Here we summarize only the most basic properties relevant to the present discussion:  The group has 
$24$ elements.  This includes $12$ elements that correspond to the $12$ proper rotations that take a regular tetrahedron into coincidence with itself, with choices of Euler 
angles that are less than $2 \pi$.  The remaining $12$ elements are the first set times an element called $R$ that corresponds to a $2 \pi$ rotation.  As we indicated earlier, $T'$ 
has ${\bf 1}$, ${\bf 2}$ and ${\bf 3}$-dimensional representations, that we specify more precisely below. For odd-dimensional representations, $R$ acts trivially and the action of the
group $T'$ is not distinguishable from that of the tetrahedral group $T$.   For the even-dimensional representations, however, $R$ acts non-trivially; this reflects the fact that 
$T'$ is a subgroup of SU(2) and that spinors flip sign under a rotation by $2 \pi$.

The complete list of $T'$ representations is as follows: there is a trivial singlet, $\bf 1^0$, two non-trivial singlets, $\bf 1^\pm$, three doublets, ${\bf 2^0}$ and ${\bf 2^\pm}$, and one
one triplet, $\bf 3$.  The different singlet and doublet representations are distinguished by how they transform under a $Z_3$ subgroup, generated by the group element
called $g_9$ in Ref.~\cite{Aranda:2000tm}.   This is indicated by the triality superscript; when we multiply representations, trialities add under addition modulo three.  Keeping this in 
mind, the rules for multiplying representations are then specified by
\begin{equation}
\begin{split}
\bd{1\otimes R} &= \bd{R\otimes 1} \text{     for any rep $\bd R$},
\\ \bd{2\otimes 2}&= \bd{3\oplus1},
\\ \bd{2\otimes 3}&= \bd{3\otimes2} = \bd{2^0\oplus 2^+ \oplus 2^-},
\\ \bd{3\otimes 3}&= \bd{3\oplus3 \oplus 1^0 \oplus 1^+ \oplus 1^-}.
\end{split}
\end{equation}

As we indicated in the Introduction, the models of interest are based on the flavor group $G_F = T' \times Z_3$, which includes a $Z_3$ subgroup that rotates
all first-generation matter fields by a phase.  We now identify that subgroup. In the models of Ref.~\cite{Aranda:2000tm}, the first two generations are assigned to 
the ${\bf 2^0}$ representation\footnote{This choice is motivated by the cancelation of discrete gauge anomalies.  See Ref.~\cite{Aranda:2000tm} for details.}, in which
the element $g_9$ is given by
\begin{equation}
\text g_9(\bd 2^0)=\begin{pmatrix} \eta^2 & 0 \\ 0&\eta \end{pmatrix},
\end{equation}
where $\eta \equiv e^{2\pi i/3}$.    However, the matter fields may also transform under the $Z_3$ factor that commutes with $T'$.  We represent charge
assignments under this $Z_3$ by an additional triality index $0, \,+ \text{ and }-$, corresponding to the phase rotations $1$, $\eta$ and $\eta^2$.   The diagonal
subgroup of the $Z_3$ subgroup generated by $g_9$ and the $Z_3$ factor that commutes with $T'$ is the intermediate symmetry that we desire; we call this 
subgroup $Z_3^D$.   If we assign the first two generations to the rep ${\bf  2^{0-}}$, then the action of $Z_3^D$ is through powers of the product 
\begin{equation}
\label{eq:Z_3^Dtrans}
\text g_9(\bd 2^0) \cd \eta^2=\begin{pmatrix} \eta & 0\\ 0&1 \end{pmatrix} \,\, ,
\end{equation}
which provides the desired first generation phase rotation.

Assigning the three generations of matter fields to the $T' \x Z_3$ reps $\bd 2^{0-} \oplus \bd 1^{00}$ yields the following transformation properties of the Yukawa matrices:
\begin{equation}
\label{eq:Yukawatrans}
Y_{U,D,E} \sim \begin{pmatrix} [\bd{3^- \oplus 1^{0-}}] & [\bd 2^{0+}] \\ [\bd 2^{0+}] & [\bd 1^{00}]\end{pmatrix}.
\end{equation}
The models of interest include a set of flavon fields, $A_{ab}$, $\phi_{ab}$ and $S_{ab}$, which transform as $\bd{1^{0-}}$,  $\bf{2^{0+}}$ and ${\bf 3^-}$, respectively.  When the $T' \x Z_3$ symmetry is broken to $Z_3^D$,  the doublet and triplet flavons acquire the VEVs
\begin{equation}
\label{eq:U(2)symbreaking}
\frac{\bk{\phi}}{M_F} \sim \begin{pmatrix}0 \\ \epsilon\end{pmatrix}, \quad \frac{\bk{S}}{M_F} \sim\begin{pmatrix}0 & 0 \\ 0 & \epsilon\end{pmatrix} \,\, ,
\end{equation}
where we use $\sim$ when we omit possible order one factors.  This is the most general pattern of non-vanishing entries that is consistent with the unbroken $Z_3^D$ 
symmetry defined by Eq.~(\ref{eq:Z_3^Dtrans}).   Yukawa couplings involving first-generation fields are generated only after the $Z_3^D$ symmetry is broken at a lower
scale; in analogy to the U(2) models of Ref.~\cite{Barbieri:1995uv,Barbieri:1996ww}, it is assumed that this is accomplished solely through the vev of the flavon $A_{ab}$,
\begin{equation}
\frac{\bk{A}}{M_F} \sim \begin{pmatrix}0&\epsilon'\\-\epsilon'&0\end{pmatrix},
\end{equation}
where $\epsilon'<\epsilon$. 
This sequential breaking $T' \x Z_3 \xrightarrow{\epsilon} Z_3^D \xrightarrow{\epsilon '} nothing$ yields a Yukawa texture for the up quarks, down quarks and leptons of the form
\begin{align}
\label{eq:yukawasymbreaking}
Y_{U,D,E} \sim \begin{pmatrix}0&\epsilon' \,\, &0\\-\epsilon'&\epsilon \,\, &\epsilon \\ 0&\epsilon \,\, &1\end{pmatrix},
\end{align}
where we've suppressed $\mathcal O(1)$ operator coefficients. 

The forms of the Yukawa matrices obtained in Eq.~(\ref{eq:yukawasymbreaking}) are inadequate, given the known differences between the up-, down- and 
charged-lepton masses.  The top quark Yukawa coupling is of order one, while the all others are substantially smaller, suggesting an additional overall suppression factor is 
desirable in $Y_D$ and $Y_E$.   Moreover, the hierarchy of quark masses is more extreme in the up-quark sector than in the down; for example, the quark mass
ratios renormalized at the supersymmetric grand unified scale are given approximately by~\cite{Ramond:1993kv}
\begin{equation}
\label{eq:masshierarchy}
 m_d :: m_s:: m_b = \lambda^4:: \lambda^2::1 \quad \text{while}\quad m_u :: m_c:: m_t = \lambda^8:: \lambda^4::1,
\end{equation}
where $\lambda\approx 0.22$ is the Cabibbo angle. This suggest that an additional suppression in the $1$-$2$ block of $Y_U$ is also desirable.  We call these suppression factors 
$\rho$ and $\xi$, which modify the textures of Eq.~(\ref{eq:yukawasymbreaking}) as follows:
 \begin{align}
\label{eq:yukawasuppresions}
Y_U& \sim \begin{pmatrix} 0 & \epsilon'\rho \,\, & 0 \\ -\epsilon'\rho & \epsilon \,\rho \,\, & \epsilon \\ 0 & \epsilon \,\, &1 \end{pmatrix},
\quad Y_D \sim \begin{pmatrix} 0 & \epsilon'  \,\, & 0 \\ -\epsilon' & \epsilon \,\, & \epsilon \\ 0 & \epsilon \,\, &1 \end{pmatrix}\xi,
\quad Y_E \sim \begin{pmatrix} 0 & \epsilon'  \,\, & 0 \\ -\epsilon' & \epsilon \,\, & \epsilon \\ 0 & \epsilon \,\, &1 \end{pmatrix}\xi.
\end{align} 
Clearly, the smallness of $\rho$ and $\xi$ does not follow directly from the assumed flavor symmetry breaking, but requires additional symmetries and/or dynamics.   In the 
U(2) models of Refs.~\cite{Barbieri:1995uv,Barbieri:1996ww} and the $T'$ models of Refs.~\cite{Aranda:1999kc,Aranda:2000tm}, $\xi$ is assumed to arise
from mixing in the Higgs sector of the theory, while the origin of $\rho$ is understood in terms of a grand unified embedding.  Flavon charge assignments under the
unified gauge group can cause Yukawa entries to arise at higher order in $1/M_F$ than they would otherwise.  In the non-supersymmetric $T'$ models that we
discuss in Sec.~\ref{sec:models}, we will neither have an extended Higgs sector nor a grand unified embedding; we will, however, show how $\rho$ and $\xi$ may
arise simply by a small extension of the flavor symmetry.

All other differences between $Y_U$, $Y_D$ and $Y_E$ can now be accommodated by the choice of the undetermined ${\cal O}(1)$ operator coefficients, identified according to naive dimensional analysis.   We generally require these to be between $1/3$ and $3$ in magnitude; the precise range is a matter of taste, but our choice is consistent with the assumptions of Refs.~\cite{Aranda:1999kc,Aranda:2000tm}.  Variations in the operator coefficients are then sufficient, for example, to account for differences 
between $Y_D$ and $Y_E$ that are attributed to group theoretic factors of $3$ in grand unified theories~\cite{georgi}.  We parameterize the Yukawa matrices in terms of 
coefficients $u_i$, $d_i$ and $\ell_i$ as follows:
\begin{align}
\label{eq:yukawatextures}
Y_U&= \begin{pmatrix} 0 & u_1\epsilon'\rho & 0 \\ -u_1\epsilon'\rho & u_2\epsilon\rho & u_3\epsilon \\ 0 & u_4\epsilon & u_5\end{pmatrix},
\quad Y_D=\begin{pmatrix} 0 & d_1\epsilon' & 0 \\ -d_1\epsilon' & d_2\epsilon & d_3\epsilon \\ 0 & d_4\epsilon & d_5\end{pmatrix}\xi,
\quad Y_E=\begin{pmatrix} 0 & \ell_1\epsilon' & 0 \\ -\ell_1\epsilon' & \ell_2\epsilon & \ell_3\epsilon \\ 0 & \ell_4\epsilon & \ell_5\end{pmatrix}\xi.
\end{align}
These forms will be used to define the Yukawa matrices at the flavor scale $M_F$ in the numerical study presented in the following section.

\section{Numerical analysis} \label{sec:numerical}

We numerically evolve the Yukawa matrices in Eq.~(\ref{eq:yukawatextures}), using the one-loop, non-supersymmetric renormalization group equations (RGEs).  The flavor 
scale $M_F$ is taken to be variable, while the scale of observable energies is chosen to be the mass of the $Z$ boson, $m_Z$.   We omit all weak-scale threshold corrections.  
The RGEs are given by~\cite{reg}
\begin{align}
\frac{dg_i}{dt}  &=\frac{b_i^{\text{SM}}}{16\pi^2} \, g_i^3, \\
\frac{d Y_U}{dt} &=\frac{1}{16\pi^2}\left(-\sum_i c_i^{\text{SM}}g_i^2+\frac{3}{2} \, Y_U Y_U^\da-\frac{3}{2}\, Y_D Y_D^\da+Y_2(S)\right) Y_U \,\,\, ,
\\ \frac{d Y_D}{dt} &=\frac{1}{16\pi^2}\left(-\sum_i c_i'^{\text{SM}}g_i^2+\frac{3}{2} \, Y_D Y_D^\da-\frac{3}{2} \, Y_U Y_U^\da+Y_2(S)\right) Y_D \,\,\, ,
\\ \frac{d Y_E}{dt} &=\frac{1}{16\pi^2}\left(-\sum_i c_i''^{\text{SM}}g_i^2+\frac{3}{2} \, Y_E Y_E^\da+Y_2(S)\right)Y_E,
\end{align}
where
\begin{equation}
Y_2(S) = \Tr[3 \,Y_U Y_U^\da+3\, Y_D Y_D^\da+Y_E Y_E^\da] \,\,\,.
\end{equation}
Here, the $g_i$ are the gauge couplings, $Y_U$, $Y_D$ and $Y_E$ are the Yukawa matrices, and
$t=\ln\mu$ is the log of the renormalization scale.  The SU(5) normalization of $g_1$ is assumed.  In the absence of 
supersymmetry~\cite{reg},
\begin{equation}
b_i^{\text{SM}} = \begin{pmatrix} \frac{41}{10}, & -\frac{19}{6}, & -7\end{pmatrix} \,\,\, ,
\end{equation}
and
\begin{equation}
\\ c_i^{\text{SM}} = \begin{pmatrix} \frac{17}{20}, & \frac{9}{4}, & 8\end{pmatrix}, 
\quad c_i'^{\text{SM}} = \begin{pmatrix} \frac{1}{4}, & \frac{9}{4}, & 8\end{pmatrix},
\quad c_i''^{\text{SM}} = \begin{pmatrix} \frac{9}{4}, & \frac{9}{4}, & 0\end{pmatrix} \,\,.
\end{equation}
The $\overline{{\rm MS}}$ gauge couplings are chosen to satisfy the boundary conditions
\begin{equation}
\begin{split}
\label{eq:gaugecouplings}
\alpha_1^{-1}(m_Z)&=59.01 \,\,\, ,
\\ \alpha_2^{-1}(m_Z)&=29.59 \,\,\, ,
\\ \alpha_3^{-1}(m_Z)&=8.44 \,\,\, ,
\end{split}
\end{equation} 
where $\alpha_i=g_i^2/4\pi$.  These were computed using the values of $\alpha_{\text{EM}} = e^2/4 \pi=127.950$ and $\sin^2\hat\theta_W=0.23129$ renormalized at $m_Z$~\cite{pdg} as 
well as
\begin{equation}
 e=g_Y\cos\hat\theta_W=g_2\sin\hat\theta_W  \,\,\,\,\, \mbox{ and } \,\,\,\,\, g_1 = \sqrt{5/3} \, g_Y \,\,\, ,
\end{equation}
where the latter equation converts the standard model hypercharge gauge coupling to SU(5) normalization~\cite{g1renorm}.  The QCD coupling is given directly in 
Ref.~\cite{pdg}.

At the flavor scale $M_F$, the Yukawa matrices are given by Eq.~(\ref{eq:yukawatextures}).  For a given numerical choice of symmetry-breaking parameters and operator 
coefficients, the Yukawa matrices are run down to the scale $m_Z$ and diagonalized.  In addition to the nine fermion mass eigenvalues, three CKM mixing angles can be
compared to experimental data.  (In this work, we do not consider the CKM phase, which is not constrained by the flavor symmetry.)  Equivalently, we take the predictions 
of the theory to consist of the nine fermion masses and the magnitudes of the three CKM elements, $V_{us}$, $V_{ub}$ and $V_{cb}$.

\begin{figure}[t]
  \begin{center}
    \includegraphics[width=0.6\textwidth]{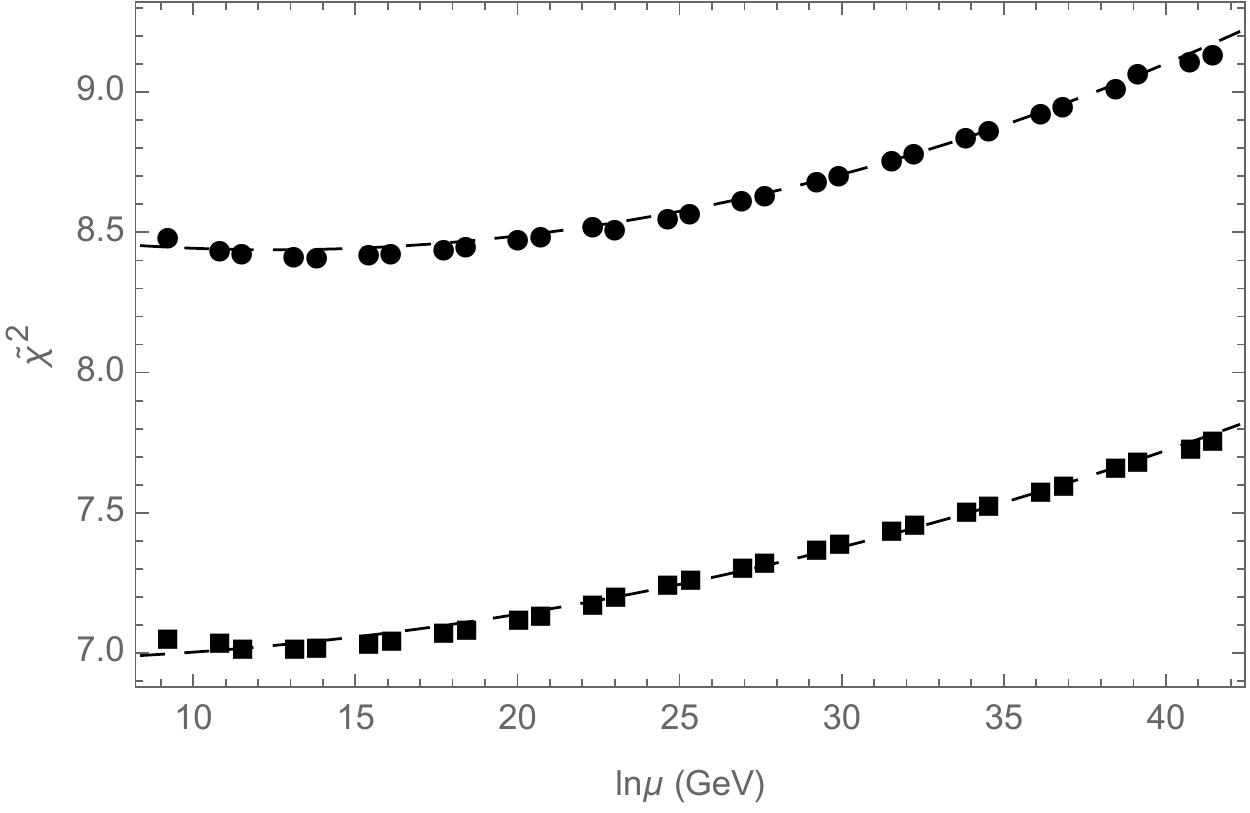}
    \caption{Minimum $\widetilde{\chi}^2$ values as a function of $M_F$, for two different model assumptions.}
      \label{fig:muvschisq}
  \end{center}
\end{figure}

To optimize the choice of parameters and operator coefficients for a given choice of flavor scale $M_F$, we follow the approach of Ref.~\cite{Aranda:2000tm} and 
minimize the function
\begin{equation}
\begin{aligned}
\label{eq:chisq}
\widetilde{\chi}^2 &= \sum_{i=1}^9 \left(\frac{m_i^{th}-m_i^{exp}}{\Delta m_i^{exp}}\right)^2+\left(\frac{|V_{us}^{th}|-|V_{us}^{exp}|}{\Delta V_{us}^{exp}}\right)^2+\left(\frac{|V_{ub}^{th}|-|V_{ub}^{exp}|}{\Delta V_{ub}^{exp}}\right)^2+\left(\frac{|V_{cb}^{th}|-|V_{cb}^{exp}|}{\Delta V_{cb}^{exp}}\right)^2
\\ &+\sum_{i=1}^5 \left(\frac{\ln |u_i|}{\ln 3}\right)^2+\sum_{i=1}^5 \left(\frac{\ln|d_i|}{\ln 3}\right)^2+\sum_{i=1}^5 \left(\frac{\ln|\ell_i|}{\ln 3}\right)^2 \,\,\, .
\end{aligned}
\end{equation}
Here, the quantities with the superscript $th$ refer to the predictions of the theory, obtained as we have described previously.  The quantities with the superscript $exp$ refer
to the experimental data, taken from Ref.~\cite{pdg}, and written as $X \pm \Delta X$, where the second term is the experimental uncertainty.   Since we've omitted two-loop 
corrections and threshold effects, we take this uncertainty into account in the same way as Ref.~\cite{Aranda:2000tm}: we inflate experimental error bars to $1$\% of the 
central value if the experimental error is smaller than this.    The terms involving ratios of logarithms in Eq.~(\ref{eq:chisq}) ensure that the operator coefficients
remain near unity~\cite{Aranda:2000tm}.  

\begin{table}[b]
\begin{center}
\caption{Fit parameters and observables for $M_F=10^6$ GeV with $\chi^2=7.021$.  In this example, the operator
corresponding to $u_4$ is absent from the theory.  All masses are given in GeV. (Note that $m_t$ is the $\overline{MS}$ 
mass, not the pole mass.)}
\vspace{2mm}
\label{tbl:res1}
\begin{tabular}{ccc} \hline\hline
\multicolumn{1}{c}{} &
\multicolumn{1}{c}{Best Fit Parameters} &
\multicolumn{1}{c}{}\\
\hline
\multicolumn{1}{c}{} &
\multicolumn{1}{c}{$\epsilon=0.182, \; \epsilon'=0.005,\; \rho=0.029,\; \xi=0.014$} &
\multicolumn{1}{c}{}\\
\hline
$u_{1}=1.131$  & $d_{1}=1.162$  & $\ell_{1}=0.651$ \\
$u_{2}=0.921$  & $d_{2}=-0.631$ & $\ell_{2}=-0.710$ \\
$u_{3}=-0.575$ & $d_{3}=1.024$  & $\ell_{3}=-1.242$ \\
$u_{4}=0$ (fixed)  & $d_{4}=2.375$  & $\ell_{4}=-1.244$ \\
$u_{5}=0.628$ & $d_{5}=-0.931$ & $\ell_{5}=-0.637$  \\
\hline \hline
\multicolumn{1}{c}{Observable}&
\multicolumn{1}{c}{Expt. Value~\cite{pdg}}&
\multicolumn{1}{c}{Fit Value}\\
\hline
$m_{u}$    & $(2.3 \pm 0.6)\x 10^{-3}$       & $1.4\x 10^{-3}$ \\
$m_{c}$    & $1.275   \pm 0.025$        & 1.277\\
$m_{t}$    & $160      \pm 4.5$          & 160.1 \\
$m_{d}$    & $(4.8   \pm 0.4)\x 10^{-3}$       & $4.18\x 10^{-3}$ \\
$m_{s}$    & $(9.5 \pm 0.5)\x 10^{-2}$        & $9.84 \x 10^{-2}$ \\ 
$m_{b}$    & $4.18     \pm 0.03$         & 4.18 \\
$m_{e}$    & $(5.11 \pm 1\%)\x 10^{-4}$ & $5.11\x10^{-4}$ \\ 
$m_{\mu}$  & $0.106    \pm 1\%$      & 0.106 \\
$m_{\tau}$ & $1.78     \pm 1\%$       & 1.78 \\
$|V_{us}|$ & $0.225 \pm 1\%$    & 0.226 \\ 
$|V_{ub}|$ & $(3.55 \pm 0.15)\x10^{-3}$      & $3.58\x10^{-3}$ \\ 
$|V_{cb}|$ & $(4.14 \pm 0.12)\x10^{-2}$       & $4.13 \x 10^{-2}$ \\
\hline\hline
\end{tabular}
\end{center}
\end{table}
We have called the function we minimize $\widetilde{\chi}^2$ to make clear that it differs from the conventional $\chi^2$ function one would define in a simple least-squares fit.  In our theory,
a choice of parameters that gives a very good match to all the experimental central values but includes an operator coefficient that is, for example,  $17.3$, would be in wild conflict 
with the assumption that we have a valid effective field theory description. Hence,  the additional terms in Eq.~(\ref{eq:chisq}) are necessary to constrain the operator coefficients independent of the 
experimental data.  Since the $u_i$, $d_i$ and $\ell_i$ are not treated as free parameters, we might expect qualitatively that a good fit will have a $\widetilde{\chi}^2 \approx 8$, corresponding to $12$ 
pieces of experimental data minus $4$ unconstrained parameters ($\epsilon$, $\epsilon'$, $\rho$ and $\xi$).  We will see that this is consistent with our numerical results.

\begin{table}[t]
\begin{center}
\caption{Fit parameters and observables for $M_F=10^{18}$ GeV with $\chi^2=7.762$. In this example, the operator
corresponding to $u_4$ is absent from the theory.  All masses are given in GeV.  (Note that $m_t$ is the $\overline{MS}$ 
mass, not the pole mass.)}
\vspace{2mm}
\label{tbl:res2}
\begin{tabular}{ccc} \hline\hline
\multicolumn{1}{c}{} &
\multicolumn{1}{c}{Best Fit Parameters} &
\multicolumn{1}{c}{}\\
\hline
\multicolumn{1}{c}{} &
\multicolumn{1}{c}{$\epsilon=0.131, \; \epsilon'=0.004,\; \rho=0.02,\; \xi=0.011$} &
\multicolumn{1}{c}{}\\
\hline
$u_{1}=1.005$  & $d_{1}=1.005$  & $\ell_{1}=0.847$ \\
$u_{2}=1.01$  & $d_{2}=-0.64$ & $\ell_{2}=-0.633$ \\
$u_{3}=-0.458$ & $d_{3}=1.024$  & $\ell_{3}=-1.193$ \\
$u_{4}=0$ (fixed)  & $d_{4}=2.397$  & $\ell_{4}=-1.199$ \\
$u_{5}=0.369$ & $d_{5}=-0.676$ & $\ell_{5}=-0.847$  \\
\hline \hline
\multicolumn{1}{c}{Observable}&
\multicolumn{1}{c}{Expt. Value~\cite{pdg}}&
\multicolumn{1}{c}{Fit Value}\\
\hline
$m_{u}$    & $(2.3 \pm 0.6)\x 10^{-3}$       & $1.4\x 10^{-3}$ \\
$m_{c}$    & $1.275   \pm 0.025$        & 1.277\\
$m_{t}$    & $160      \pm 4.5$          & 160.4 \\
$m_{d}$    & $(4.8   \pm 0.4)\x 10^{-3}$       & $4.2\x 10^{-3}$ \\
$m_{s}$    & $(9.5 \pm 0.5)\x 10^{-2}$        & $9.8 \x 10^{-2}$ \\ 
$m_{b}$    & $4.18     \pm 0.03$         & 4.18 \\
$m_{e}$    & $(5.11 \pm 1\%)\x 10^{-4}$ & $5.11\x10^{-4}$ \\ 
$m_{\mu}$  & $0.106    \pm 1\%$      & 0.106 \\
$m_{\tau}$ & $1.78     \pm 1\%$       & 1.78 \\
$|V_{us}|$ & $0.225 \pm 1\%$    & 0.226 \\ 
$|V_{ub}|$ & $(3.55 \pm 0.15)\x10^{-3}$      & $3.58\x10^{-3}$ \\ 
$|V_{cb}|$ & $(4.14 \pm 0.12)\x10^{-2}$       & $4.13 \x 10^{-2}$ \\
\hline\hline
\end{tabular}
\end{center}
\end{table}

A plot of $\widetilde{\chi}^2$ as a function of the flavor scale $M_F$ is shown in Fig. \ref{fig:muvschisq}.   The two curves in this figure correspond to the cases were the coefficient $u_4$ is allowed to
float, or is fixed to zero.  (In the latter case, the sum over the $u_i$ in the second line of Eq.~(\ref{eq:chisq}) omits $i=4$.) These cases are motivated by two variants of the Yukawa textures that may arise in explicit models, as we show in Sec.~\ref{sec:models}.  Over the entire range of $M_F$ we find good fits with $\widetilde{\chi}^2 \approx 8$, but with clear and monotonic improvement in $\widetilde{\chi}^2$ towards smaller values of $M_F$.   In addition, the case where the operator corresponding to $u_4$ is absent from the theory ({\em i.e.}, where $u_4$ is fixed to zero), which we will see corresponds to more minimal model-building assumptions, provides a better description of the  data than the case where it is present.  We present two examples of our results in Tables~\ref{tbl:res1} and \ref{tbl:res2}, for $M_F=10^{6}$~GeV and $10^{18}$~GeV, respectively, both in the case where $u_4=0$.   
The first choice corresponds to a flavor scale of the same order as the lower bounds from flavor-changing neutral current processes, as we discuss further in the next section, while the second is of the same order as 
the Planck scale.   Interestingly, the latter demonstrates that the model is consistent with the possibility that their are only two important  physical scales in nature, the weak and the Planck scales (with flavor associated with the latter) so that no additional
hierarchies or fine-tuning need to be considered.

\section{Direct lower bounds on the flavor scale} \label{sec:fcnc}

Our results in Fig~\ref{fig:muvschisq} indicate that typical $T'$ Yukawa textures provide a good description of charged fermion masses and CKM angles over
a wide range of $M_F$, but with a preference for values closer to the TeV scale than to the Planck scale.  The lowest possible values of 
$M_F$ are separately constrained by flavor-changing-neutral-current (FCNC) processes that receive contributions from heavy flavor-sector fields.   In this 
section, we provide some estimates of the lower bounds on $M_F$ following from $K^0 - \overline K^0$, $D^0 - \ol D^0$, $B^0 - \ol B^0$ and $B_s^0 - \ol B_s^0$ 
mixing.   In addition, we give the branching fractions predicted for the largest flavor-changing neutral meson decays, which also violate lepton flavor.

The new physics contributions to the FCNC processes of interest come from flavon exchange, or more precisely, the exchange of the 
physical fluctuations about the flavon vevs. We identify these as follows:
\begin{equation}
\phi = \begin{pmatrix} \varphi_1 \vspace{2mm} \\ \epsilon\, M_F+\varphi_2\end{pmatrix}, 
\quad S_{ab} = \begin{pmatrix}  \tilde{S}_{11} & \tilde{S}_{12} \vspace{2mm} \\ \tilde{S}_{12} & \epsilon \, M_F + \tilde{S}_{22}\end{pmatrix}, 
 \quad A_{ab} = \begin{pmatrix} 0 & \epsilon' \, M_F+ \tilde{A} \vspace{2mm} \\ -\epsilon' \, M_F-\tilde{A} & 0\end{pmatrix} \,\,\,,
 \label{eq:fluc}
\end{equation}
where  the $\varphi_i$, the $\tilde{S}_{ij}$ and $\tilde{A}$ are complex scalar fields.  The couplings to standard model fermions originate from the same operators
that gave us the Yukawa couplings.  As an example, let us consider the origin of $\Delta S=2$ operators, where $S$ here refers to strangeness.
We focus on the largest flavor-changing effects, ones that are present even in the absence of a rotation from the gauge to mass eigenstate basis.
Let $\Psi$ be a three-component column vector with the elements $d$, $s$ and $b$.   Then the flavon-quark-anti-quark vertex in the down sector 
follows from
\begin{equation}
\mathcal L \supset - \frac{v}{\sqrt{2}}( \ol\Psi_L Y_D \Psi_R+ \text{h.c.}) \,\,\, ,
\end{equation}
where we have set the standard model Higgs field to its vev $v/\sqrt{2}$, where $v=246$~GeV, and where
\begin{equation}
Y_D= \left(
\begin{array}{c|c}
S_{ab}/M_F+A_{ab}/M_F  & \phi/M_F \\ \hline
\phi/M_F  & 1
\end{array}
\right) \, \xi \,\,\, ,
\end{equation}
with the flavons $S$, $A$ and $\phi$ given by Eq.~(\ref{eq:fluc}), and $\xi$ is the dimensionless suppression factor defined earlier.  (We provide an
origin for $\xi$ and $\rho$ in the next section.)  The flavon couplings involving fermions of the first two generations only are given by
\begin{equation}
d_1\frac{v\, \xi}{\sqrt2 M_F} (\ol d_L \tilde{A} \,s_R-\ol s_L \tilde{A} \,d_R)-d_2\frac{v\, \xi}{\sqrt2 M_F} (\ol d_L \tilde{S}_{12} \, s_R
+ \ol s_L \tilde{S}_{12}\, d_R) + \text{h.c.}  \,\,\,.
\end{equation}
Four-fermion operators are obtained by integrating out the heavy fields.  It follows that the $\Delta S=2$ operator that contributes to the 
$K^0 - \ol K^0$ mass splitting is 
\begin{equation}
{\cal O}_{\Delta S=2} = -\left(\frac{d_1^2}{m_{\tilde{A}}^2}+\frac{d_2^2}{m_{{\tilde{S}}_{12}}^2}\right) \frac{v^2 \xi^2}{2 M_F^2} \, [\ol d_L s_R \ol d_R s_L],
\label{eq:ourds2}
\end{equation}
where the $d_i$ are the same order one coefficients defined in Eq.~(\ref{eq:yukawatextures}).  As the flavon masses are not known exactly, we assume that they are of the same 
order as the symmetry-breaking scale associated with the given flavon; in the present example,
\begin{equation}
m_{{\tilde{S}}_{12}} \sim \epsilon \, M_F  \quad \mbox{and} \quad m_{{\tilde{A}}} \sim \epsilon' \, M_F \,\,\, .
\end{equation}
Moreover, we pick numerical values of $\epsilon$, $\epsilon'$, $\rho$ and $\xi$ that are characteristic of the values found in our global fits for $M_F$
below $\sim 1000$~TeV:
\begin{equation}
\epsilon \sim 0.1, \quad \xi \sim 0.03, \quad \rho\sim 0.02 \,\,\, .
\end{equation}
\begin{table}[t]
\begin{center}
\begin{tabular}{ccc} \hline \hline
\multicolumn{1}{c}{Mass Splitting} &
\multicolumn{1}{c}{Operator} & 
\multicolumn{1}{c}{$M_F$ Lower Bound}\\
\hline
 $K^{0}-\ol{K^{0}}$          &  $-d_{2}^{2}\frac{1}{m^{2}_{{\tilde{S}}_{12}}}\frac{v^2 \xi^2}{2 M_F^2} \ol{d}_{L}s_{R}\ol{d}_{R}s_{L}$  & 85 TeV \\
                              &                                                                                                                                    &       \\
  $B^{0}-\ol{B^{0}}$          &  $-d_{3}d_{4}\frac{1}{m^{2}_{{\varphi}_{1}}}\frac{v^2 \xi^2}{2 M_F^2} \ol{d}_{L}b_{R}\ol{d}_{R}b_{L}$                            & 22 TeV \\
                              &                                                                                                                                    &       \\
 $B^{0}_{s}-\ol{B^{0}_{s}}$  &  $-d_{3}d_{4}\frac{1}{m^{2}_{\varphi_{2}}}\frac{v^2 \xi^2}{2 M_F^2} \ol{b}_{L}s_{R}\ol{b}_{R}s_{L}$                            & 14 TeV\\
                              &                                                                                                                                    &       \\
$D^{0}-\ol{D^{0}}$          &  $-u_{2}^{2}\frac{1}{m^{2}_{{\tilde{S}}_{12}}} \frac{v^2\rho^{2}}{2 M_F^2}    \ol{u}_{L}c_{R}\ol{u}_{R}c_{L}$ & 14 TeV \vspace{0.3em} \\ 
\hline\hline
\end{tabular}
\end{center}
\caption{Lower bounds on the flavor scale.  See the text for definitions of our notation.}
\label{tbl:lowerbound}
\end{table}
We set all order one coefficients equal to one.   With these assumptions, the new physics contribution to the neutral pseudoscalar meson mass splittings, $\Delta m$,
may be expressed as a function of the scale $M_F$.   In general, given a $\Delta F=2$ interaction of the form $c \, {\cal O}$, where $c$ is the operator coefficient and
$F$ represents either strange (S), charm (C) or bottom (B), the mass splitting is given by
\begin{equation}
\label{eqn:masssplitting}
\Delta m= \frac{c}{m_{P^0}}\big|\langle P^0| \mathcal O | \ol P^0 \rangle\big| \,\,\, ,
\end{equation}
where $P^0 \, (\ol P^0)$ is the pseudoscalar meson (anti-meson) in question, and the states are relativistically normalized.  For an operator of the form
\begin{equation}
\label{eqn:operator}
\mathcal O=\frac{1}{4}[\ol h^\alpha (1-\gamma^5)\ell ^\alpha][\ol h^\beta(1+\gamma_5)\ell^\beta] \,\,\, ,
\end{equation} 
where $h,\,\ell$ represent the heavy (light) quark flavors and $\alpha,\,\beta$ are color indices, the matrix element in Eq.~(\ref{eqn:masssplitting}) is given by~\cite{k&d}
\begin{equation}
\label{eq:k&d}
\bk{P^0 | \mathcal O|P^0} = \frac{1}{2}B_{P^0}\frac{m^4_{P^0}f_{P^0}^2}{(m_h+m_\ell)^2},
\end{equation}
in the case where $P^0=K^0$ or $D^0$.  Here, $B_{P^0}$ is the bag parameter, $m_{P^0}$ and $f_{P^0}$ are the mass and decay constants of the meson and 
$m_\ell, \, m_h$ are the masses of the quarks that make up the meson.   For $P^0=B^0$ or $B_s^0$, the matrix element is given by~\cite{b&bs}
\begin{equation}
\label{eq:b&bs}
\bk{P^0 | \mathcal O|P^0} = \frac{1}{2}B_{P^0} f_{P^0}^2\ m^2_{P^0}\left[\left(\frac{m_{P^0}}{m_h+m_\ell}\right)^2+\frac{1}{6}\right]  \,\,\,.
\end{equation}
As computed on the lattice, the bag parameter in Eq.~(\ref{eq:k&d}) is defined by the expression as shown~\cite{k&d}, omitting the additional term proportional to $1/6$ that is retained in Eq.~(\ref{eq:b&bs}); in 
the case where $P^0=K^0$ or $D^0$, the effect of this term is negligible.  All masses and mass splittings were obtained from the Review of Particle Properties~\cite{pdg}, all decay constants were 
obtained from Ref.~\cite{decays}, the bag parameters for $\Delta S=2$ and $\Delta C=2$ were obtained from Ref.~\cite{k&d}, and the bag parameters for $\Delta B=2$ were obtained from Ref.~\cite{b&bs}.  To estimate the lower
bound on $M_F$, we assume that the experimentally observed mass splittings are consistent with the standard model predictions and require that the new physics contributions not exceed the 
current $2 \sigma$ experimental uncertainty.   Such an approach is sufficient for an estimate given the theoretical uncertainties involved in determining the new physics contribution itself.   Our results are shown in Table~\ref{tbl:lowerbound}. As one might expect, we obtain the tightest bound from the $K^{0}-\bar{K^{0}}$ mass splitting, which 
requires $M_F \gtrsim 85$ TeV. 
\begin{table} [t]
\begin{center}
\begin{tabular}{ccccc}
\hline\hline
\multicolumn{1}{c}{Decays} &
\multicolumn{1}{c}{BF (Ref.~\cite{pdg})}&
\multicolumn{1}{c}{ Operator } & 
\multicolumn{1}{c}{$M_F$ Lower Bound}\quad&
\multicolumn{1}{c}{BF ($M_F=85$~TeV)} \\ \hline
&&&&\\
$\;\; K_L^0 \to \ol \mu e \quad$ & $<4.7\x 10^{-12}$ & \quad $-d_{2}\ell_2  \frac{1}{m^{2}_{{\tilde{S}}_{12}}} \frac{v^2 \xi^2}{2 M_F^2} \ol{e}_{L}\mu_{R}\ol{s}_{R}d_{L}$ \quad& 9.8 TeV &  $1.5 \x 10^{-19}$
\\ &&&& \\
$\;\;B^0 \to \ol \tau e \quad$ & $<2.8 \x 10^{-5}$  & $-d_{4}\ell_3  \frac{1}{ m^{2}_{{\varphi_1}}} \frac{v^2 \xi^2}{2 M_F^2}\ol{e}_{L}\tau_{R}\ol{d}_{R}b_{L}$ &
0.62 TeV &  $2.3 \x 10^{-22}$
\\ &&&& \\
$\;\;B_s^0 \to \ol \tau \mu \quad $ &  --- &  $-d_{3}\ell_4 \frac{1}{ m^{2}_{{\varphi_2}}} \frac{v^2 \xi^2}{2 M_F^2} \ol{s}_{L}b_{R}\ol{\mu}_{R}\tau_{L}$ & --- &
$3.2 \x 10^{-22}$
\\ &&&& 
\\ \hline\hline
\end{tabular}
\end{center}
\caption{\small{Lower bound on $M_F$ for the largest flavor-changing decays.  The predicted branching fraction for $M_F$ set equal to the $K^0$-$\bar K^0$ mixing
bound is also shown.}}
\label{tbl:branchingfractions}
\end{table}

Flavon exchange between quarks and leptons can also lead to flavor-changing neutral meson decays.  We again focus on operators that are flavor-changing 
in the absence of a rotation of the fields from the gauge to mass eigenstate basis.  The largest effects are shown in Table~\ref{tbl:branchingfractions}.  The relevant
operators are of the form ${\cal O}_{qde}^{ijkn} \equiv (\overline{\ell}_i e_j)(\overline{d}_k q_n)$, in the notation of Ref.~\cite{bfconstraints}; in the same
reference, bounds on the operator coefficients are conveniently summarized.  We translate these into bounds on the scale $M_F$ which, as can be seen from
Table~\ref{tbl:branchingfractions}, are much weaker that those coming from the pseudoscalar meson mass splittings.   Therefore, we also show the predicted
branching fractions with $M_F$ set equal to our lower bound from $K^0$-$\overline{K^0}$ mixing.   It is clear that the predicted branching fractions
are far below the experimental bounds and unlikely to have observable consequences.   Note that we have only considered CP conserving processes and it is
generally known that bounds on CP violation in the neutral kaon system tends to give a better bound on the scale of new physics by about an order of magnitude
compared to the CP-conserving FCNC bounds.  Given the smallness of these branching fractions, this fact does not change our qualitative conclusions, so we
do not pursue that issue further.

\section{Nonsupersymmetric models} \label{sec:models}

In the renormalization group analysis of Sec.~\ref{sec:numerical}, the Yukawa matrices $Y_i$ are defined by
\begin{equation}
{\cal L}_{m} = \frac{v}{\sqrt{2}} \overline{\psi^i_L} Y_i \psi^i_R + \mbox{ h.c.} \,\,\,\ ,
\end{equation}
where $i=U$, $D$ or $E$ and generation indices are suppressed.  In order to replicate the Yukawa textures of the 
supersymmetric models of Refs.~\cite{Aranda:1999kc,Aranda:2000tm}, we assign the right-handed fermions of the three 
generations to the $T' \times Z_3$ representations ${\bf 2^{0-}} \oplus {\bf 1^{00}}$.  Hence, for example, we would assign the 
first two generations of the charge-$2/3$ quarks according to $(u^c_L , c^c_L) \sim (u_R , c_R) \sim {\bf 2^{0-}}$, where the 
superscript ``c'' refers to charge conjugation;  since $\overline{\psi} = i {\psi^c}^T \gamma^0 \gamma^2$, this is equivalent to 
specifying the transformation properties of the Dirac adjoints $(\overline{u_L} , \overline{c_L})$.   
We then identify the following transformation properties for the various blocks of 
the $Y_i$,
\begin{equation}
Y_{U, \,D,\, E} \sim \left(\begin{array}{cc} [{\bf 3^-} \oplus {\bf 1^{0-}}] &  [{\bf 2^{0\,+}}] \\
{[{\bf 2^{0\,+}}]} & [{\bf 1^{00}}] \end{array} \right)\,\, ,
\label{eq:ycharges}
\end{equation}
{\em i.e.}, Eq.~(\ref{eq:yukawatextures}) (or Eq.~(4.1) in Ref.~\cite{Aranda:2000tm}), which omits any additional symmetries that 
may be needed to explain the suppression factors $\rho$ and $\xi$.   As in the supersymmetric model, the transformation properties given in Eq.~(\ref{eq:ycharges}) determine the allowed flavon couplings.  However, in the supersymmetric case, 
Eq.~(\ref{eq:ycharges}) dictates the form of terms in the superpotential, which is required to be a holomorphic function of the 
superfields. The absence of this constraint in the nonsupersymmetric case could lead, in principle, 
to additional flavon couplings that are not present in the supersymmetric theory.  However, we see that as far as the 
$\phi$, $S$ and $A$ flavons are concerned, this is not the case:  each has a nontrivial $Z_3$ charge, which prevents 
new flavon couplings at the same order that involve the complex conjugates of these fields. 

In the supersymmetric theories of Refs.~\cite{Aranda:1999kc,Aranda:2000tm}, the additional suppression factors associated
with the parameters $\rho$ and $\xi$ required the introduction of additional fields and symmetries.
For example, in the simplest unified $T' \times Z_3$ model of Refs.~\cite{Aranda:1999kc,Aranda:2000tm}, SU(5) charge assignments
of the flavon fields are responsible for forbidding the coupling of the  $A$ and $S$ flavons in $Y_U$ at lowest order in $1/M_F$. However, these couplings emerge via higher-order operators that involve a flavor-singlet, SU(5) adjoint field $\Sigma \sim {\bf 24}$, just as in earlier models based on
U(2) flavor symmetry~\cite{Barbieri:1996ww}. The suppression associated with the parameter $\xi$, on the other
hand, was assumed to arise via mixing in the Higgs sector, a reasonable possibility since
supersymmetric models require more than one Higgs doublet.

Here we will also achieve the additional suppression factors by means of additional fields
and symmetries. However, the additional symmetry will be much smaller than the product
of supersymmetry and a grand unified gauge group. (The latter, of course, would not be
appropriate for the non-supersymmetric case where the gauge couplings do not unify.) We
will simply assume an additional $Z_3$ factor, so that the flavor group is $G_F^{new} = T' \times (Z_3)^2$
Defining one of the elements of the new $Z_3$ factor as $\omega = \exp(2\, i\, \pi /3)$, the only standard
model fields that transform nontrivially under this symmetry are
\begin{equation}
H \rightarrow \omega \, H  \,\,\,\,\,  \mbox{ and }  \,\,\,\,\, t_R \rightarrow \omega \,t_R \,\,\, ,
\label{eq:z3c}
\end{equation}
where $H$ is the standard model Higgs field and $t_R$ is the right-handed top quark. In the standard
model, $H$ couples to $Y_D$ and $Y_E$, while $\sigma^2 H^*$ couples to $Y_U$. Hence, when the new $Z_3$
symmetry is unbroken, the assignments in Eq.~(\ref{eq:z3c}) forbid $Y_D$ and $Y_E$ entirely, as well as
the first two columns of $Y_U$. How one proceeds with the model building depends on the desired relative sizes 
of $\epsilon$, $\epsilon'$, $\rho$ and $\xi$.  For example, for some choices of $M_F$, it is possible to
find numerical results that are consistent with the simple possibility $\epsilon \sim \rho \sim \xi$, up to order one factors.  
In this case, we assume the symmetry-breaking pattern
\begin{equation}
T' \times (Z_3)^2 \stackrel{\epsilon}{\longrightarrow} Z^D_3 \stackrel{\epsilon'}{\longrightarrow} \mbox{ nothing} \,\,\, ,
\end{equation}
where the intermediate $Z^D_3$ factor is exactly the same one as in the original theory, that transforms
all first generation fields by a phase; in this case, the new $Z_3$ symmetry is broken at
the first step in the symmetry-breaking chain.  We introduce two new flavon fields
\begin{equation}
\rho \rightarrow \omega^2 \, \rho \,\,\,\,\, \mbox{ and } \,\,\,\,\, \tilde{\phi} \rightarrow \omega\, \tilde{\phi} \,\,\, ,
\end{equation}
where $\tilde{\phi}$ transforms like $\phi\sim{\bf 2^{0+}}$ under the original flavor group. With the assumed 
symmetry breaking pattern, the $\rho$ field and one component of the $\tilde{\phi}$ doublet can develop vevs of
order $\epsilon \, M_F$. The $Z_3$ charges of these fields now allow us to rebuild our otherwise forbidden
Yukawa matrices as follows:

({\em i}.) For $Y_D$ and $Y_E$, we may generate matrices proportional to the standard form if we
replace $H$ by $H\,\rho$; it follows that $\langle\rho\rangle/M_F$ is identified with the suppression factor $\xi$, which we
now predict to be of order $\epsilon$, up to an order one factor.  One might worry that we could obtain a lower-order contribution 
from operators that don't involve $\rho$, but involve $\tilde{\phi}^*$ instead, which also transforms under the
new $Z_3$ factor as $\tilde{\phi}^* \rightarrow \omega^2 \tilde{\phi}^*$. However, this does not occur 
since $\tilde{\phi}^* \sim {\bf 2^{0-}}$ under the original flavor symmetry, which is not one of the representations 
that leads to a lowest order coupling. On the other hand, the product   $\rho^* \tilde{\phi}$ does couple at the 
same order as $\rho\, \phi$; however, this additional contribution does nothing to the form of the resulting Yukawa
textures beyond a redefinition of the order one coefficients.

({\em ii}.) For $Y_U$, the two-by-two block associated with the flavons $A$ and $S$ can now be
recovered via operators involving $\rho^* A$ and  $\rho^* S$. Hence, the parameter we called $\rho$ 
previously is now predicted to be of the same order as $\xi$.  In an analogous way, the 3-1 and 3-2 entries of $Y_U$ can 
couple to the product $\rho^*\phi$, but this transforms in the same way as $\tilde{\phi}$, which may couple at lower-order. 
Hence the canonical $Y_U$ texture with an additional suppression in only the upper-left two-by-two block is obtained.
Note that we could simply omit $\tilde{\phi}$ from the theory and ignore the corresponding entries in
$Y_U$; this leads to an alternative texture in which $u_4=0$ in Eq.~(\ref{eq:yukawatextures}), neglecting corrections
from higher-order operators.  This was the alternative possibility considered in Sec.~\ref{sec:numerical}.
It is worth noting that in the case where the $\tilde{\phi}$ is omitted from the theory, there is no longer a necessary connection between
the scale of the additional $Z_3$ breaking and the scale of the $T'$ doublet vev, $\epsilon M_F$.  In this case, we could vary 
this additional scale independently so that $\rho$ and $\xi$ are still comparable, but intermediate in size between $\epsilon$ and
$\epsilon'$.  This construction would be compatible with the numerical results in Tables~\ref{tbl:res1} and \ref{tbl:res2}.

In summary, we have provided an existence proof that the textures considered in our numerical analysis may arise in 
a relatively simple way in a non-supersymmetric framework.

\section{Conclusions} \label{sec:conc}

In this paper, we have reconsidered models of flavor based on the non-Abelian discrete flavor group $T'$ that were proposed in
Ref.~\cite{Aranda:1999kc,Aranda:2000tm}.  We have relaxed two assumptions made in these studies, that the models are 
supersymmetric and that the scale of the flavor sector is around the scale of supersymmetric grand unification.  Our numerical study 
found that $T'$ models without supersymmetry provide a viable description of charged fermion masses and CKM angles for a 
range of values of the flavor scale $M_F$.  We find that identification of $M_F$ with the reduced Planck scale is a viable 
possibility, consistent with a simple picture in which no new physics appears between the weak and gravitational scales.  However,
we also find that our fits improve monotonically as $M_F$ is lowered toward the lower bound dictated by the constraints from 
flavor-changing-neutral-current processes.  In the case where $M_F$ is as low as possible, we identified the largest
flavor-changing neutral current effects that result from the exchange of heavy flavor-sector fields; these could provide
indirect probes of the model.  We then showed how the form of the Yukawa textures that we studied, which were the same as, or closely related to,
those described in Ref.~\cite{Aranda:1999kc,Aranda:2000tm}, can nonetheless arise in a non-supersymmetric framework, where there is only
a single Higgs doublet field and where the interactions do not originate from a superpotential, a holomorphic function of the fields.   The 
models we described are arguably simpler than their supersymmetric counterparts; in the non-supersymmetric case, we needed only to extend 
the original flavor-group by a $Z_3$ factor to obtain the desired Yukawa textures shown in Eq.~(\ref{eq:yukawatextures}), while avoiding the 
well-known complications that come with a grand unified Higgs sector.   Extending the present study to include the neutrino sector is more 
model dependent, but would be interesting for future work.
  
\begin{acknowledgments}  
This work was supported by the NSF under Grant PHY-1519644. In addition, S.V. thanks the William \& Mary Research
Experience for Undergraduates (REU) program for support under NSF Grant PHY-1359364.
\end{acknowledgments}


\end{document}